\documentclass{revtex4}%
\begin{document}
\title{DIAMAGNETISM OF 2D-FERMIONS IN THE STRONG NONHOMOGENEOUS STATIC
MAGNETIC FIELD \( {\bf B}  =B( 0, 0, 1/cosh^{2}( \frac{x-x_{0}
}{ \delta }))\) : gas magnetization, static magnetic susceptibility, chemical potential and gas compressibility.}
 \author{M. Hud\'{a}k \\ Stierova 23, SK - 040 23 Kosice, Slovak Republik \\
and \\
O. Hud\'{a}k \\ Department of Aviation Technical Studies, Faculty of Aerodynamics, Technical University Kosice, \\ Rampova 7, SK - 040 01 Kosice, Slovak Republik}

\begin{abstract}
We study diamagnetism of a gas of fermions moving in a nonhomogeneous magnetic
field \( {\bf B}  = B( 0, 0, 1/cosh^{2}( \frac{x-x_{0}}{ \delta })). \)
The gas magnetization, the static magnetic susceptibility, the chemical
potential and the gas compressibility are discussed and compared with the uniform
field case. General need to study dynamics of electrons in different types of magnetic
fields follows from a large number of experimental situations in which its
understanding enables physicists to obtain new information.
\end{abstract}
\maketitle
\section*{PACS Numbers:}
\begin{description}
\item[71.45] Electron gas - electron states (condensed matter) ,
\item[75.20] Diamagnetism ,
\end{description}

\section*{}
Recently, using results of our exact description of the spinless
fermion motion in a nonhomogeneous magnetic field \( {\bf B}  =
B( 0, 0, 1/cosh^{2}( \frac{x-x_{0}}{ \delta })) \), we studied ground
state energy properties of a gas of spinless fermions moving in this field.
For densities lower than some critical value, \(  \nu <  \nu_{c}(B, \delta ), \)
the corresponding total energy is lower than that of the uniform field state,
\cite{[OH]}. However, physical properties of a gas of spinless two-dimensional
(2d) -fermions moving in this field were not studied in \cite{[OH]}. It is
the aim of this paper to fill in this gap.

Energy of a gas of free spinless fermions moving in a plane in a uniform field
perpendicular to this plane is larger or equal to its total energy
\(  E_{T}( 0, \nu ) = 2 \pi tN \nu^{2} \) in the zero field:
\begin{equation}
\label{eq:1}
\Delta E_{h}(n) \equiv E_{T}( B, \nu )-E_{T}(0, \nu ) = 2 \pi tN( \nu_{n+1} -
\nu )( \nu - \nu_{n}),
\end{equation}
where \( \nu_{n+1} \geq \nu > \nu_{n},  \nu_{n} \equiv n. \frac{ \Phi}{ \Phi_{0}},
n= 0, 1, ...;  \Phi_{0} \) is a unit of the magnetic flux, \( \Phi \equiv Ba^{2},
t \equiv \frac{ \hbar^{2} }{2ma^{2}}; \) \( \nu \) is the number
density of the gas, \( \nu \equiv \frac{N_{f}}{N} \), \( N_{f} \) is the total
number of fermions, \( Na^{2} \) is the area of the square with the side
length \( L=a \sqrt{N}, \) to which the motion is bounded, m is the fermion mass.
Here \( a \) is a characteristic length of the system, its value is of
the order of a lattice constat value.
The energy degeneracy occurs in (\ref{eq:1}) only if a number of Landau
levels is completely filled (e.i. if for some n \( \nu_{n}= \nu ). \)
A nonhomogeneity of the field introduced by a
local field intensity decrease leads to competition of two
tendencies: a decrease of the single
fermion energy level due to decreased value of the field intensity
and a decrease of the every energy level degeneracy due to 
larger spacing between centers of neighboring orbits within
the region of smaller fields. Spectrum of 2d Bloch electrons
in a periodic magnetic field was studied in \cite{[BBR]}, where the magnetic
unit cell is assumed to be commensurate with the lattice unit cell.
Their work is extension of previous studies of free electrons in
periodic magnetic field \cite{[N]} to the lattice case. In our work
we concentrate our attention on the nonperiodic modulated field.

Let us consider motion of a spinless fermion gas bounded to the square LxL in a
nonhomogeneous static magnetic field perpendicular to this plane. We neglect
the lattice periodic potential influence on the gas energy 
spectrum in this paper. We consider in more details the limit in which
nonhomogeneity disappears and a uniform field appears.
In difference to \cite{[BBR]} and \cite{[N]} we do not consider a periodic
magnetic field.  Recently, \cite{[4]}, an exact
description of motion of the quantum spinless fermion in a nonhomogeneous
magnetic field described by the vector potential \( {\bf A} =
( 0, B \delta tanh( \frac{x-x_{0}}{ \delta}), 0) \) was found.
Let us now describe those results from \cite{[4]} which are relevant for our
further calculations in this paper. In the case of motion of a quantum
spinless fermion in a nonhomogeneous magnetic field described by the vector
potential \( {\bf A} = ( 0, B \delta tanh( \frac{x-x_{0}}{ \delta}), 0) \)
the energy spectrum of the motion in the x-direction is splitted, see in
\cite{[4]}, into a discrete and a continuous parts for general values of
the field B and of the nonhomogeneity parameter \( \delta. \)
We take \( x_{0}=0 \) in the following, thus the
field has its maximum intensity at \( x=0. \) Let us consider
the limit of strong fields \( ( F \equiv 2 \pi \frac{ \Phi^{'}}{ \Phi} >>
1/2, \) where  \( \Phi^{'} \equiv B \delta^{2}). \) In this limit it is
sufficient to take into account the lowest energy levels of the spectrum.
The eigenvalues of the energy corresponding to
this part of the spectrum are given by, see in \cite{[4]} :
\[ E_{n}(p)=
\frac{p_{y}^{2}}{2m}(1- \frac{F^{2}}{((\frac{1}{4}+F^2)^{\frac{1}{2}}-((1/2)+n)
)^{2}})+ \]
\[ ( \frac{ \hbar^{2}}{2m \delta^{2}})(F^{2}-((\frac{1}{4}+F^{2})^{\frac{1}{2}}
-( \frac{1}{2}+n))^{2}, \]
where \( n= 0, 1,...[n_{max}], \) here [n] denotes an integer part of a real
number n, \( p_{y} \) is the y-momentum. Let us define \( P \equiv
\frac{ \mid p_{y} \mid \delta }{ \hbar } \). The number \( n_{max} \) is defined
by:
\[ n_{max} = ( \frac{1}{4}+F^{2})^{ \frac{1}{2} }-(1/2)-( \mid P \mid F)^
{ \frac{1}{2}}, \]
for given values of P and F.

For \( F^{2} >> \frac{1}{4} \) and for small quantum numbers n, \( n= 0, 1, 2,
 ..., \) the energy \( E_{n}(p_{y}) \) expanded into series of 1/F powers takes
the form :
\[ E_{n}(p_{y}) \approx \hbar \omega (n+ \frac{1}{2})- \\
( \frac{ \hbar^{2}}{2m \delta^{2}} )
((n+ \frac{1}{2})^{2}+ \frac{1}{4})- \frac{p_{y}^{2}}{mF}(n+ \frac{1}{2})+ \]
\[( \frac{ \hbar^{2}}{8mF \delta^{2}})(n+ \frac{1}{2})+ O( \frac{1}{F^{3}}). \]
where \( \omega \equiv \frac{Bc}{em} \) is the cyclotron frequency.
The energy levels become degenerated in the
limit of strong but modulated fields if the energy expansion above
is restricted to the first two terms, which are of the \( F^{1} \) and \( F^{0} \)
orders respectively.
The largest value of the third term in this expansion is negligible with respect
to the second term
\[  max( \frac{p^{2}_{y}}{mF}(n+ \frac{1}{2})) <<
( \frac{ \hbar^{2}}{2mF \delta^{2}})(n+ \frac{1}{2}). \]
if we take into account that there exists a natural cut-off
for \( p_{y} \) momenta, \( max( \mid p_{y} \mid ) = \frac{ \pi \hbar }{ a}, \)
due to the underlying crystal and if we assume that the field intensity B
satisfies the inequality:
\[ \frac{ \Phi}{ \Phi_{0} } >>  8 \pi^{2}, \]
where \( \Phi \equiv B.a^{2} .\)
The degeneracy of the n-th level appears due to the lost of the energy
dependence on \( p_{y} \) momentum.
These levels are, \cite{[4]}, modified Landau levels with energies in the form
\begin{equation}
\label{eq:2}
E_{n} = \hbar \omega (n + \frac{1}{2}) - \frac{ \hbar^{2}}{2m \delta^{2} }
[(n + \frac{1}{2})^{2} + \frac{1}{4}] + O(1/F),
\end{equation}
where  \[ n = 0, 1, ...  <<  n_{m}; n_{m} \approx F. \]
Note that
\[ \frac{ \hbar^{2}}{2m \delta^{2}} = 4t( \frac{L}{2 \delta})^{2}/N. \]
Every energy level \( E_{n} \) is degenerated within considered
approximation, its degeneracy \( D_{n} \) is found to be:
\begin{equation}
\label{eq:3}
D_{n} = D_{L} \frac{tanh( \frac{L}{2 \delta })}{ \frac{L}{2 \delta }}.
\end{equation}
Here the characteristic length L and the nonhomogeneity parameter \( \delta \)
satisfy the inequality:
\[ tanh(L/2 \delta ) < (1- \frac{2}{F}(n +  \frac{1}{2})). \]
The Landau level degeneracy is \( D_{L} \equiv \frac{Bea^{2}}{hc} N \)
as in the case of the uniform field. The form of the degeneracy
\( D_{n} \) given above holds for all orders of F. The large F
expansion in (\ref{eq:2}) limits its validity to the region of system parameters
given by the inequality below (\ref{eq:3}) which follows from
the usual, \cite{[5]}, boundary conditions: periodicity in the y-direction
perpendicular to the x-axis and limits on the position of
the orbit center in the x-direction to the region \( < -L/2, +L/2 >. \)
The orbit center x-coordinate \( x_{c} \) is given, \cite{[4]}, by
\[ tanh(x_{c}/ \delta ) = ( \frac{-p_{y} \delta }{ \hbar})/F. \]

The ground state energy \( E_{T}(B, \delta , \nu ) \) for spinless fermion gas
with density \( \nu \) in the limit of strong but nonhomogeneous fields
specified by \( B, \delta \) given as difference between the nonhomogeneous
field state and the zero field state energy is found to be:

\begin{equation}
\label{eq:4}
\Delta  E_{nh}(n) \equiv  E_{T}(B, \delta , \nu ) - E_{T} (0, \nu ) =
 2 \pi t N [( \nu -  \nu_{n}
\frac{tanh( \frac{L}{2 \delta } )}{ \frac{L}{2 \delta } } )
( \nu_{n+1} \frac{ tanh( \frac{L}{2 \delta } ) }{ \frac{L}{2 \delta } } -  \nu ) +
\end{equation}
\[ ( 1 - \frac{tanh( \frac{L}{2 \delta } ) }{ \frac{L}{2 \delta } })( \nu (2 \nu_{n}+
 \nu_{1}) - \frac{tanh( \frac{L}{2 \delta } ) }{ \frac{L}{2 \delta } }
\nu_{n+1} \nu_{n} )] - \]
\[ - \frac{ta^{2}}{ \delta^{2} } N [ \nu (n^{2} + n + \frac{1}{2}) -  \nu_{n}
( \frac{2n^{2}}{3} + n + \frac{1}{3} )]. \]

The total energy difference (\ref{eq:4}) is found assuming that there are
n levels \( 0, 1, ..., n-1 \) filled
and that the n-th level is filled partially. The gas
density \( \nu \) in (\ref{eq:4}) is limited by the following inequalities:
\begin{equation}
\label{eq:4'}
\nu_{n+1} \frac{ tanh( \frac{L}{2 \delta } ) }{ \frac{L}{2 \delta } } \geq \nu >
\nu_{n} \frac{ tanh( \frac{L}{2 \delta } ) }{ \frac{L}{2 \delta } },
\end{equation}
\[ \nu_{n} \equiv n \Phi / \Phi_{0}. \]
The uniform field result (\ref{eq:1})
follows from (\ref{eq:4}) and (\ref{eq:4'}) in the
limit \( \delta \longrightarrow \infty \) keeping values of
all the other system parameters constant.

Let us now calculate some of physical characteristics of our gas
of spinless fermions in the considered nonhomogeneous magnetic field
at zero temperature. Then let us compare these characteristics
( magnetization, static magnetic susceptibility, chemical potential
and compressibility ) with the corresponding characteristics in the
uniform field case.
The comparison enable us to describe influence of nonhomogeneity
of the magnetic field in the large intensity and nonhomogeneity limit
on a gas of fermions.

The magnetization is found to be given by
\begin{equation}
\label{eq:5}
m_{z}(B, \delta ) \equiv \frac{1}{N} \frac{ \partial E_{T}( B, \delta, \nu )}
{ \partial B} = ( \frac{a^{2}}{ \Phi_{0} } ).
[ 2 \pi t ( \nu (2n + 1) - \frac{tanh( \frac{L}{2 \delta } ) }{ \frac{L}{2 \delta } }
(n+1)2n \nu_{1} )] +
\end{equation}
\[  \frac{ta^{2}}{ \delta^{2} } n ( \frac{2n^{2}}{3} + n + \frac{1}{3} )]. \]

We see from (\ref{eq:5}) that when the nonhomogeneity parameter \( \delta \)
increases its value to infinity we obtain the uniform field result. When
this parameter is finite then the magnetization increases its value ( its
absolute value decreases).

The static magnetic susceptibility \( \chi_{zz}(B, \delta ) \) is found to be
given by:
\begin{equation}
\label{eq:6}
\chi_{zz}(B, \delta ) \equiv \frac{ \partial m_{z}}{ \partial B} =
( \frac{a^{2}}{ \Phi_{0} } )^{2}.
[- 4 \pi t n.( n+1 ) \frac{tanh( \frac{L}{2 \delta } )}{ \frac{L}{2 \delta } } ]
\end{equation}

We see that the susceptibility becomes less diamagnetic than in the uniform
field case due to presence of the field nonhomogeneity where the field
intensity is decreased, and thus also where the density of states is lower.

The chemical potential difference between the nonhomogeneous field state
and the zero field state \( \Delta \mu \) is given by
\begin{equation}
\label{eq:7}
\Delta \mu = \frac{1}{N} . \frac{ \partial \Delta  E_{nh}(n)}{ \partial \nu}
= 2 \pi t [( \nu_{1} (2n+1) - 2 \nu ] - \frac{ta^{2}}{ \delta^{2} }
[ n^{2} + n + \frac{1}{2}].
\end{equation}
The single fermion energy shift down due to presence of nonhomogeneity
decreases the chemical potential and is described in (\ref{eq:7}) by the second
term. The first term has the same form as in the uniform field case. Note,
however, that the density of particles, \( \nu \), is in the nonhomogeneous
field case bounded from above and below by limiting values which are
dependent on the nonhomogeneity of the field, see (\ref{eq:4'}). In the
uniform field case there exists symmetry when filling a given energy
level \( ( \nu \rightarrow \nu_{n+1}^{-} ) \) and when emptiing the same level
\( ( \nu \rightarrow \nu_{n}^{+}) \):
\[ \mu( \nu \rightarrow \nu_{n}^{+} )= - \mu( \nu \rightarrow \nu_{n+1}^{-}). \]
This symmetry is lost whenever the nonhomogeneity parameter is finite.

The difference of the inverse compressibility \( \Delta \frac{1}{ \kappa} \)
of the gas in the nonhomogeneous field state and the zero field state
is given by
\begin{equation}
\label{eq:8}
\Delta \frac{1}{ \kappa} \equiv - \frac{ \partial \mu }{ \partial \nu}
= 4 \pi t
\end{equation}
From (\ref{eq:8}) it follows that the inverse compressibility is not
affected by the presence of the nonhomogeneity. Thus decrease of the chemical
potential with increase of the gas density is the same in the uniform field case and
in the nonhomogeneous field case.

Results of this paper describe some of the physical properties of
2d-fermion gas moving in our specific type of the nonhomogeneous field.
We expect that qualitatively these results describe modification of
the uniform field case due to presence of a static nonhomogeneity of general type.
General need to study dynamics of electrons in different types of magnetic
fields follows from a large number of experimental situations in which its
understanding enables physicists to obtain new information. In the solid
state physics these situations occur studying such effects as: the Hall effect;
magnetoresistivity; anomalous skin effects in a magnetic field; cyclotron
resonances in metals; magnetoplasma waves in metals and semiconductors;
quantum oscillatory effects like: the de Haas - van Alfen effect, oscillations
of the entropy, of the volume, of the specific heat, of the thermopower and of
other thermodynamic characteristics, the Shubnikov-de Haas effect,
oscillations of the surface impedance, of the Hall coefficients, of the sound
absorption coefficients and of the sound velocity, magnetoacoustic oscillations
(Pippard's geometric resonance oscillations), giant quantum oscillations of
absorption in metals, quasiclassical size effects (radiofrequency size effects,
cutt off effects of cyclotron resonance and of quantum oscillations,
oscillations on cutted off orbits, the Sondheimer effect) and quantum size
effects (mesoscopic phenomena).

\end{document}